\documentclass{IEEEtran} 
\usepackage{textcomp}
\usepackage[noadjust]{cite}

\usepackage[T1]{fontenc}
\usepackage{amsmath}
\usepackage{caption}
\usepackage{multirow}
\usepackage{bm}
\usepackage{diagbox}

\usepackage[cmintegrals]{newtxmath}
\usepackage{bm}
\usepackage{array}
\usepackage{graphicx}
\usepackage{xcolor}
\usepackage{url}

\ifCLASSINFOpdf
\else
\fi

\begin{document}

\title{Using Synthetic Data to Enhance the Accuracy of Fingerprint-based Localization: A Deep Learning Approach}

\author{\IEEEauthorblockN{Mohammad~Nabati$^1$, Hojjat~Navidan$^1$, Reza~Shahbazian{$^1$}{$^{,i}$}, Seyed~Ali~Ghorashi{$^1$,$^2$}{$^{,ii}$},
and~David~Windridge{$^3$}}
\\ \IEEEauthorblockA{{$^1$}Cognitive Telecommunication Research Group, Department of Telecommunications, Faculty of Electrical Engineering, Shahid Beheshti University, G.C., Tehran, Iran\\
{$^2$}Department of Computer Science \& Informatics, School of Architecture, Computing and Engineering, University of East London (UEL), London, UK\\
{$^3$}Department of Computer Science, School of Science and Technology, Middlesex University, London, UK\\
{$^i$}Member, IEEE, 
{$^{ii}$}Senior Member, IEEE}%

\thanks{Corresponding author: S. Ali Ghorashi (e-mail: a\_ghorashi@sbu.ac.ir).\protect\\}}

\maketitle

\begin{abstract}
Human-centered data collection is typically costly and implicates issues of  privacy. Various solutions have been proposed in the literature to reduce this cost, such as crowd-sourced  data collection, or the use of semi-supervised algorithms. However, semi-supervised algorithms require a source of unlabeled data, and crowd-sourcing methods require numbers of active participants. An alternative passive data collection modality is fingerprint-based localization. Such methods use received signal strength (RSS) or channel state information (CSI) in wireless sensor networks to localize users in indoor/outdoor environments. In this paper, we introduce a novel approach to reduce  training data collection costs in fingerprint-based localization by using synthetic data. Generative adversarial networks (GANs) are used to learn the distribution of a limited sample of collected data and, following this, to produce synthetic data that can be used to augment the real collected data in order to increase overall positioning accuracy. Experimental results on a benchmark dataset show that by applying the proposed method and using a combination of 10\% collected data and 90\% synthetic data, we can obtain essentially similar positioning accuracy to that which would be obtained by using the full set of collected data. This means that by employing GAN-generated synthetic data, we can use 90\% less real data, thereby reduce data-collection costs while achieving acceptable accuracy.
\end{abstract}

\begin{IEEEkeywords}
Synthetic Data, Generative Adversarial Networks, Deep Learning, Fingerprint Localization, Wireless Sensor Networks.
\end{IEEEkeywords}

\section{Introduction}
The rapid development of smartphones has led to increasing demand for location-based services (LBSs) based on wireless sensor networks within different areas, including academic research, industry and commerce \cite{R1,R3,R20}. Localization services such as global positioning system or global navigation satellite system are typically only available in outdoor environments, and, even there, such satellite-based methods may not provide acceptable accuracy in all outdoor environments due to non-line of sight error, fading, and shadowing effects \cite{R2}. These approaches employ ranging-based methods such as time of arrival, angle of arrival and RSS to estimate the location of the users. These methods also can be used to localize the LBSs in indoor environments. However, they do not provide acceptable accuracy.

Fingerprinting methods are used to improve the accuracy of LBSs in indoor environment \cite{R3}. They constitute a subset of localization approaches in which the signal of multiple base stations (BSs) such as WiFi, Bluetooth, ZigBee, light, and radio-frequency identification is used \cite{R2} to determine the location of a receiver. Among these wireless systems, WiFi has attracted the most interest due to its ready availability in modern smartphones and other communication devices \cite{R3}. In this paper, we use the term Access Point (AP) to refer to WiFi BS. Fingerprinting methods have two distinct phases: the offline (or training) and online (or test) phases. The mode of data collection in the offline phase depends on the localization purpose, a typical purpose being to determine a particular zone that the user is  located in. In such problems, class labels are assigned to each area, and fingerprints such as RSS or CSI are collected for each class separately (defining a classification problem). Alternatively, when an exact location estimation is required (in essence, a regression problem), during the training phase fingerprints such as RSS or CSI are collected at specific locations of the environment, known as reference points. In the online phase, the target user receives RSS or CSI from multiple BSs or APs. This information is then passed  to the system model trained during  the offline phase such that  the area or location of the user is finally estimated. In this paper, we focus on the classification problem and use RSS as the method for receiving signals from multiple APs.

Machine learning methods, and particularly deep learning approaches, have recently been used to recognize the statistical patterns of gathered datasets to train the system model in the offline phase of fingerprint-based localization \cite{R20}. A deep model consists in several layers of neural networks connected by weighted links with activation functions applied to the outputs; typically, such methods  define the state-of-the-art in classification performance across a range of domains. However, due to the parametric complexity of these models, the accuracy of deep-learning based fingerprint localization methods depends strongly on the number of samples in the training phase. Because such training data is necessarily in short supply in practical domains, there is a corresponding motivation to establish new methods that can reduce data collection costs, while reaching an accuracy comparable to that of idealized deep neural classification.

Many researchers have tried to reduce the data collection cost. Authors in \cite{R6} propose a hybrid generative-discriminative approach for handling unlabeled data using a small quantity of labeled data. In \cite{R7}, the authors propose a semi-supervised deep-learning approach to reduce the cost of collecting labeled data based on constructing a neighborhood graph (similarity matrix) from trajectory data (conventional methods for constructing a  neighborhood matrix do not consider the physical location of labeled data). The authors of \cite{R9} also propose an approach (GrassMA) that explicitly takes the location of labeled data into account. Authors in \cite{R21} use deep belief networks to update the hidden features of labeled fingerprints via the sufficient unlabeled data.

All of the methods mentioned above require "real unlabeled" data to reach an acceptable level of accuracy. Data-gathering without loss of user privacy is a challenging issue. In this paper, we propose a new method to improve the accuracy of localization while using less real data; we will thus use GANs to produce synthetic data as an extra input to the classification model. GANs have  been used for augmentation of data in classification problems within a number of  other research areas, including 
remote sensing  \cite{R13}, object classification \cite{R14}, and more generally in computer vision \cite{R15}. To the best of our knowledge, this is the first time that synthetic data generated by a GAN  has been used to improve  fingerprint-based localization.

In this paper, we use lower-case letters (e.g. $a$) to denote scalar numbers, boldface lower-case letters (e.g. $\mathbf{a}$) for vectors, capital letters (e.g. $A$) for functions and boldface capital letters (e.g. $\mathbf{A}$) to denote matrices.

The remainder of this paper is organized as follows: in section II, the conventional system model for fingerprint-based classification problem is presented. In section III, the proposed deep learning-based model to generate synthetic data is set out. Experimental results and conclusions are presented in sections IV and V, respectively.

\begin{figure}[!t]
\vspace*{-10pt}
\centering
\includegraphics[width=\columnwidth]{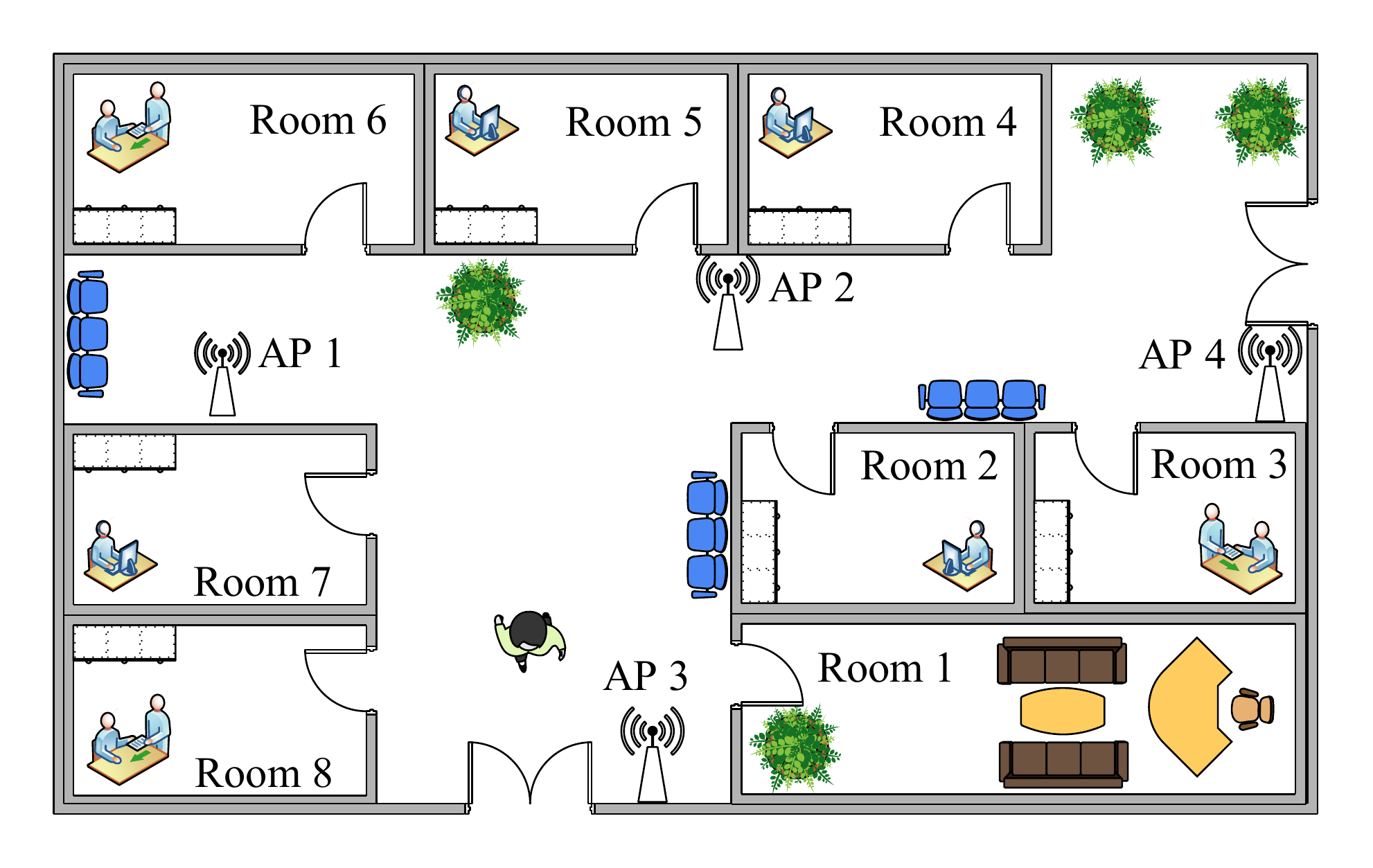}
\caption{A schematic of the indoor environment for localization.}
\label{fig:1}
\end{figure}

\section{System Model}
The schematic of a typical indoor environment classification problem is depicted in Fig. \ref{fig:1}.  Initially, labels are assigned for each area or room. Then, RSS values from multiple APs are gathered for each area such that a system model can be trained from the  collected dataset.
We use a deep neural network, $F(\mathbf{x},{\theta _c})$ to model task-related patterns with respect to the room-configuration using the  RSS values, 
where, $\mathbf{x} \in {\mathbb{R}^{\mathbf{1 \times M}}}$ is the input vector (the RSS vector in our problem), ${\theta _c}$ is the parameter of the deep model learned  during the training phase, and the $C$ represents the class output nodes. Given the multi-class nature of the classification problem, we make use of a cross-entropy or log-likelihood loss function \cite{R16}:
\begin{equation}
{\cal L}({\theta _c}) =  - \sum\limits_{i = 1}^N {\sum\limits_{j = 1}^C {{y_{ij}}\log {{\hat y}_{ij}}} } 
\label{eq:1}
\end{equation}
where ${y_{ij}} = \left\{ {1\,\,\,{\rm{if}}\,\,i \in c,\,\,0\,\,\,\,{\rm{o.w}}} \right\}$, $N$ is the total number of observations, $C$ the number of classes, ${y_{ij}}$ a  real label value and ${\hat y_{ij}}$ the predicted value arising from the trained $F(\mathbf{x},{\theta _c})$. Equation \ref{eq:1} can be written in vector form via the class summation:
\begin{equation}
{\cal L}({\theta _c}) =  - \sum\limits_{i = 1}^N {{\mathbf{y}_i}\log \mathbf{\hat y}_i^T} 
\label{eq:2}
\end{equation}
where ${\mathbf{y}_i} \in {\mathbb{B}^{1 \times C}}\,$ and $\,\mathbb{B} \in \{ 0,1\} $ i.e.  ${\mathbf{y}_i}$ is one-hot encoded  so as to be one of ${\left[ {1,{\text{ }}0,{\text{ }}0, \ldots ,{\text{ }}0} \right]_1}$, ${\left[ {0,{\text{ 1}},{\text{ }}0, \ldots ,{\text{ }}0} \right]_2}$, ..., ${\left[ {0,{\text{ }}0,{\text{ }}0, \ldots ,{\text{ 1}}} \right]_C}$.  ${\hat y_{ij}}$ is a real number between 0 and 1  predicted by the deep model. The log function favors hard selection of a single class; back-propagation is used to train the system via minimization of  the log-likelihood cost function using the   Adam \cite{R19} optimizer to obtain $F(\mathbf{x},{\theta _c})$, such that  the  maximum value of the  outputs of $C$ nodes constitutes  the class predicted for a given input vector  by the trained system model.

\section{Proposed Method}
\begin{figure*}[!t]
          \vspace*{-25pt}
	\centering
	\includegraphics[width=6.8in]{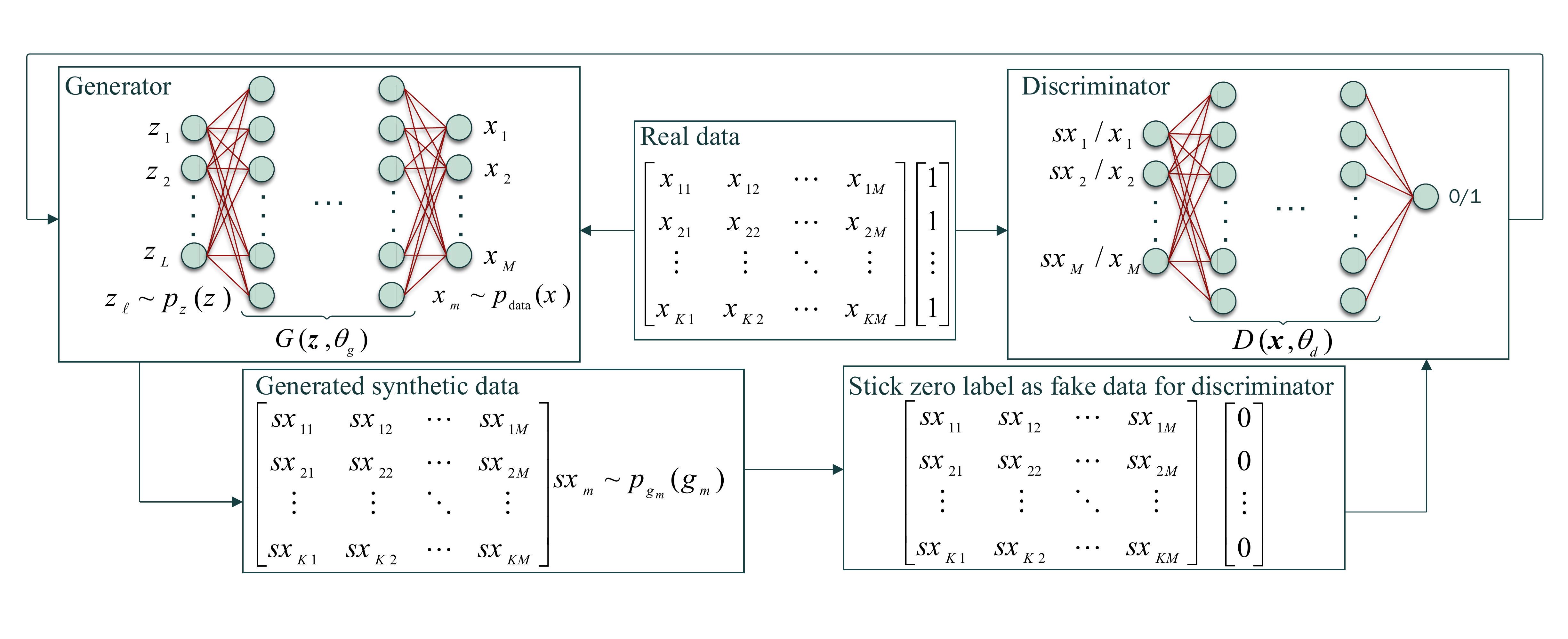}
          \centering
          \setlength{\abovecaptionskip}{-10pt}
	\caption{The broad  structure of GAN in the training phase, where $M$ is the dimension of input vector and $K$ is the number of observations.}
	\label{fig:3}
\end{figure*}
Generative Adversarial Networks, introduced by Goodfellow et al. in 2014 \cite{R10}, are a class of game-theoretic  methods used for learning the feature-distribution of a given dataset, so as to be able to parametrically-generate synthetic data with maximal similarity to the input. GANs generally consist of two distinct  parts:  a Generator and a Discriminator. The generator is responsible for learning the distribution of the training dataset and generating simulated data (via input noise) that matches the distribution of original data. The discriminator takes these data as input, and through comparison with real data, seeks to evaluate their authenticity. By continuously training these two networks together, it is hoped that a convergent point is reached in which the generator is  able to create synthetic data that sufficiently matches the distribution of real data so as to be able to  fool the discriminator. 

We shall first consider  a hypothetical dataset consisting in a problem of class discrimination for some small number of classes. Suppose a GAN-based process  produces synthetic data for each class.  Since the process for one class is the same as that of  the other classes, let the RSS dataset for this one class  be as follows:

\begin{equation}
\mathbf R_c = \left[ {\begin{array}{*{20}{c}}
	{r_{11}}&{r_{12}}& \cdots &{r_{1M}} \\ 
	{r_{21}}&{r_{22}}& \cdots &{r_{2M}} \\ 
	\vdots & \vdots & \ddots & \vdots  \\ 
	{r_{K1}}&{r_{K2}}& \cdots &{r_{KM}} 
	\end{array}} \right]
\label{eq:3}
\end{equation}
\noindent where $M$ is the number of APs in the environment, $K$ is the number of class observations, and $r_{ij}$ is the magnitude of the i'th observation from the j'th AP. Each column of the above matrix constitutes a distribution over the desired class. Therefore, we define $\mathbf{x} \in {\mathbb{R}^{\mathbf{1 \times M}}}$ such that  the goal of the generator is to map the prior noise latent variables $\mathbf{z} \in {\mathbb{R}^{\mathbf{1 \times L}}}$ to the distribution  of  $\mathbf R_c$.

The broad structure of GAN designed to achieve this is depicted in Fig. \ref{fig:3}. The process for producing the synthetic data is hence based on the  cost function:
\begin{equation}
\begin{array}{l}
\mathop {\min }\limits_G \mathop {\max }\limits_D {\mkern 1mu} {\mkern 1mu} {\cal L}(D,G),\,\,\,  \rm{where}\\
{\cal L}(D,G) = {E_{x\sim{p_{\rm{data}}}(x)}}[\log D({\mathbf{x}})] + {E_{z\sim{p_z}(z)}}[\log (1 - D(G({\mathbf{z}})))]
\end{array}
\label{eq:4}
\end{equation}
It may be seen that the cost function consists of two-parts, with the goal of the discriminator being to maximize the probability of correctly assigning labels to the real and synthetic data. Both $G$ and $D$ are differentiable functions represented by a multilayer perceptron (MLP). The Generator learns how to map the latent noise $\mathbf{z}\sim{p_z}(z)$  to the real data distribution ${\mathbf x\sim{p_{\rm{data}}}(x)}$, denoted via the $G(\mathbf{z},{\theta _g})$  structure, where ${\theta_g}$  indicates  the parameters of the MLP in the generator. Contrarily, the discriminator learns how to distinguish between real and synthetic data denoted via the  $D(\mathbf{x},{\theta _d})$ structure, where ${\theta _d}$ indicates  the parameters of the discriminator. The discriminator has a binary classification structure in its output, in which 0 and 1 indicate synthetic and real data, respectively. The cost function can be represented for the generative and discriminative models, respectively, by fixing the  other component. Therefore, the discriminator loss is defined as follows:
\begin{equation}
{\cal L}({\theta _d}) = {E_{x\sim{p_{\rm{data}}}(x)}}[\log D(\mathbf{x,}{\theta _d})] + {E_{z\sim{p_z}(z)}}[\log (1 - D(G(\mathbf{z,}{\theta _g}))]
\label{eq:5}
\end{equation}
The loss function of generator is correspondingly defined as follows:
\begin{equation}
{\cal L}({\theta _g}) = {E_{z\sim{p_z}(z)}}[\log (1 - D(G(\mathbf{z,}{\theta _g}))]
\label{eq:6}
\end{equation}
The first term of Eq. (\ref{eq:4}) vanishes during the gradient update step since  it does not effect  the generator when  fixed. 

The process for updating  ${\theta _d}$ and ${\theta _g}$ until convergence is given in full  in algorithm 1. For both the generator and discriminator, Adam optimizer is used to update the parameters. The important note is that updating the discriminator is performed in $s$ times rather than the generator. Convergence occurs when $D(\mathbf{x},{\theta _d}) = \frac{1}{2}$, meaning that the discriminator is not able to distinguish between real and synthetic data. After convergence, the generator  is  ready to produce synthetic samples for the desired class via the same prior noise distribution $\mathbf{z}\sim{p_z}(z)$.

In the next stage of the proposed fingerprint localization pipeline, this  synthetic data is combined with real data from  each class in order to augment  conventional classification via the deep learning  model  used for  training, as discussed in the previous section. Therefore, the full set of RSS ($\mathbf {FR}$) data,  which consists of both real RSS ($\mathbf {R}$)  and synthetic RSS ($\mathbf {SR}$) data for the desired class $c$, can be defined as follows:
\begin{equation}
\mathbf {FR}_c = \left( {\begin{array}{*{20}{c}}
	{\mathbf R_c} \\ 
	{\mathbf {SR}_c} 
	\end{array}} \right)
\label{eq:7}
\end{equation}
where, $\mathbf R_c \in {\mathbb{R}^{K \times M}}$, $\mathbf {SR}_c \in {\mathbb{R}^{P \times M}}$, $\mathbf {FR}_c \in {\mathbb{R}^{(K + P) \times M}}$. 

\begin{figure}[!t]
	\centering
          \vspace*{-20pt}
	\includegraphics[width=3.3in]{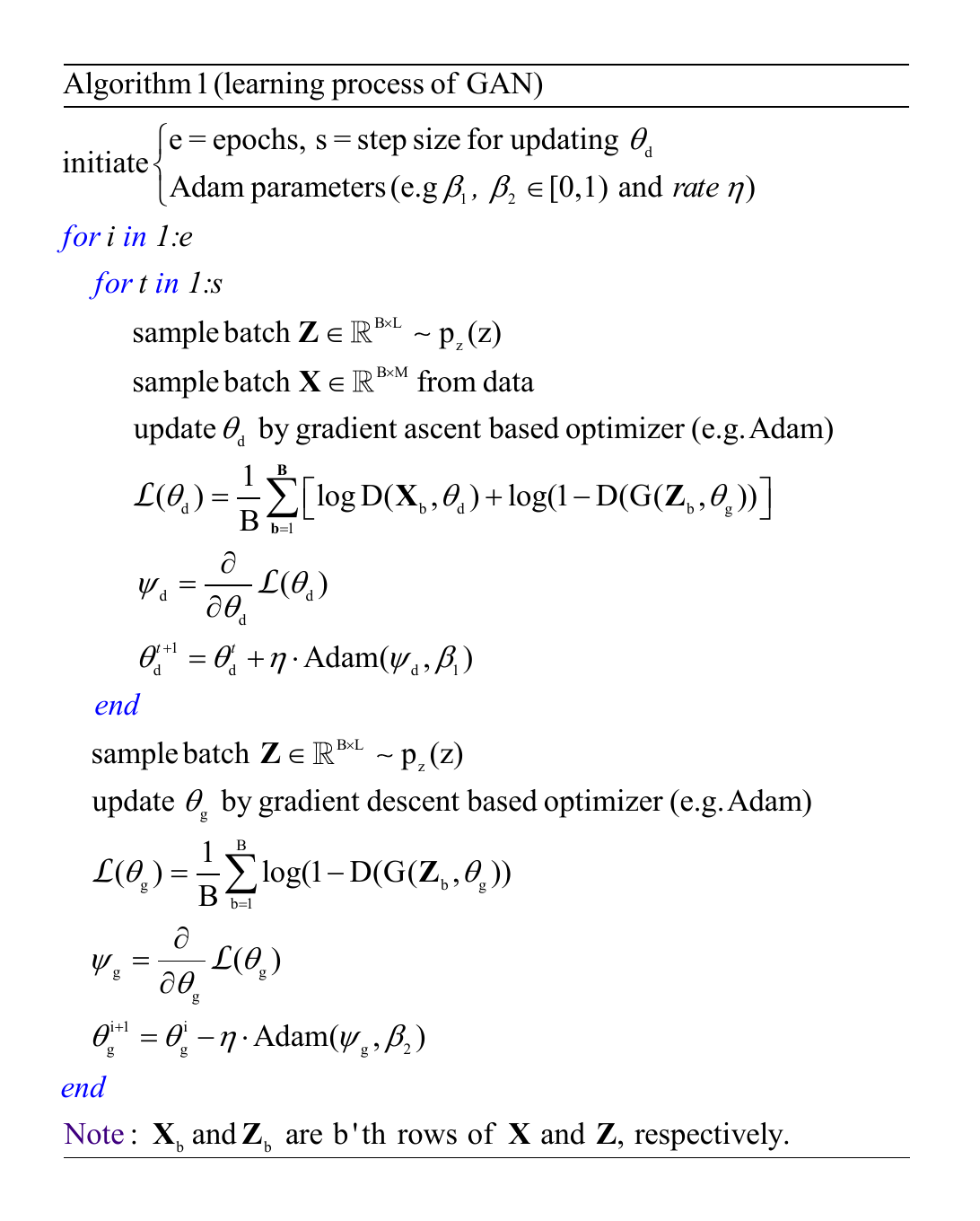}
	\caption*{}
	\label{fig:temp}
         \vspace*{-30pt}
\end{figure}
\section{Experiment Methodology and Results}
The dataset\footnote{The dataset is available in:\\\url{https://archive.ics.uci.edu/ml/datasets/Wireless+Indoor+Localization}} used in this paper is provided by Rajen Bhatt \cite{R17}, and consists of 2000 RSS samples collected from 7 APs in four different rooms (4 classes). We randomly select half of the data for training and the other half for the test phase. Both training and test datasets contain 250 data samples from each class. All of the data are standardized before presenting them to the classification model. The model used for classification is an MLP consisting of 6 densely connected layers. 
The inputs are hence RSS samples, and outputs are the class likelihoods. Both classification and GAN models have been implemented on Tensorflow 1.13 and accelerated by Geforce RTX 2060.

We perform multiple experiments to demonstrate the validity of the proposed method; in the first experiment, 10\% of the training data in each class (25  samples) is randomly selected and presented to the GAN model in order to generate synthetic data. We add the data generated by the GAN to the remaining dataset to increase the total quantity of data. The second experiment is similar to the first one, with the exception that all of the training data (250 samples) are selected. Results in terms of  test accuracy and log-likelihood loss for both experiments are presented in Table \ref{tab:1}. Accuracy is defined as $\left( {}^{{{N}_{\rm{true}}}}\!\! \diagup \!\!{}_{{{N}_{\rm{total}}}}\; \right)\times 100$, where ${{N}_{\rm{true}}}$ is the number of true predicted classes within the  test data and ${{N}_{\rm{total}}}$ is the total number of test data points (equal to 1000 in our experiments); the log-likelihood loss is defined in Equation \ref{eq:1}. As can be seen in this table, the classification model is saturated after adding 750 and 250 samples to the 10\% and 100\% of the real data such that adding the extra synthetic samples does not increase the accuracy. In order to minimize the sample bias effects in the results, the process of randomly selecting data, measuring the classification accuracy, generating synthetic data and determining  the final accuracy is carried-out several times. Each neural network model is trained and validated over 100 times with different initial model seeds. The average of the test accuracy and test loss from these runs are then reported as final results. 

\begin{figure}[!t]
          \vspace*{-6pt}
	\centering
	\includegraphics[width=2.5in]{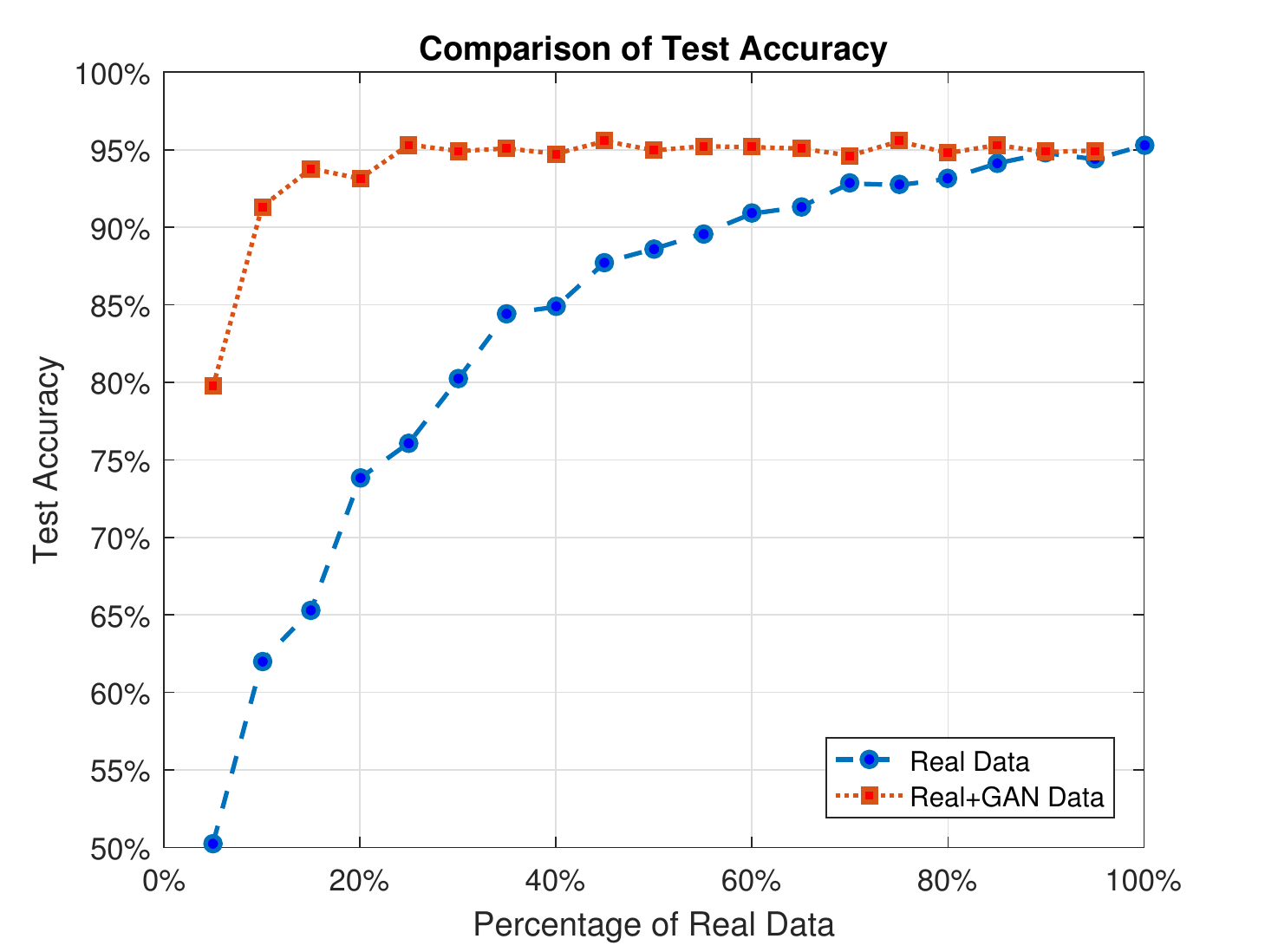}
	\caption{Comparison of classification accuracy between purely real data (blue line) and real data combined with synthetic data (red line).}
	\label{fig:4}
\end{figure}

In the final experiment, a fraction of the real data is randomly selected in order to generate synthetic data such that the total quantity of data after adding synthetic data  is equal to the quantity of original data in each class (e.g. $K$=250). As depicted in Fig. \ref{fig:4}, by carrying-over the fraction of real data used from 5\% to 100\%, we measure the test accuracy and loss in order to evaluate the effect of synthetic data on the classification accuracy. Fig. \ref{fig:4} indicates that  the test accuracy of the model is around 50\% when only a small fraction of real data is used; however, by adding synthetic data, it  increases to a value of 80\%. It can be seen that the effect of using real data after 90\% has diminished and the classification model is saturated. Therefore, as a result from Table \ref{tab:1} and Fig. \ref{fig:4}, it cannot be expected a miracle from synthetic data to significantly enhance the accuracy, especially when adequate samples of real data are available.

\begin{table}[!t]
\centering
\caption{Effect on test accuracy and log-likelihood loss of adding synthetic data samples to 10\% and 100\% of real data.}
\begin{tabular}{|c|c|c|c|c|} 
\hline
\multirow{2}{*}{\diagbox[width=12em, height = 2.3em]{\kern-0.7em \bf Synthetic Data}{\bf Real Data \kern-0.7em}} & \multicolumn{2}{c|}{\bf 10\% (25 Samples)} & \multicolumn{2}{c|}{\bf 100\% (250 Samples)} \\ \cline{2-5} 
                                 &\bf Accuracy         &\bf Log Loss     &\bf Accuracy          &\bf Log Loss         \\ \hline
\bf0                          &62.0\%                 &1.03                 &95.3\%                  &0.14         \\ \hline
\bf250                      &92.6\%                 &0.24                 &97.1\%                  &0.08         \\ \hline
\bf500                      &93.0\%                 &0.28                 &97.4\%                  &0.08          \\ \hline
\bf750                      &94.5\%                 &0.24                 &97.3\%                  &0.08          \\ \hline
\bf1000                    &94.4\%                 &0.25                 &97.2\%                  &0.08          \\ \hline
\end{tabular}
\label{tab:1}
\end{table}

\section{Conclusion}
In this paper, we have demonstrated  that synthetic data can improve the accuracy of fingerprint-based localization in a deep learning context, where the data collection process is time-consuming and costly. In  particular, we have proposed the use of a specialized GAN implementation in order to generate synthetic data to provide high-accuracy localization.  Experimental results indicate that the proposed method for classification using only 10\% of real data combined with generated synthetic data can get very close to the accuracy of a similar system using 100\%  real labeled data. This reduces the expense of data collection significantly, and encourages us to apply the concept to fields other  than fingerprint-based localization.

\normalsize

\bibliographystyle{IEEEtran}
\bibliography{ref}

\end{document}